\documentclass[a4paper]{jpconf}

\usepackage{graphicx}

\usepackage{amsmath}
\usepackage{amssymb}
\usepackage{amsfonts}

\usepackage{color}

\definecolor{light-gray}{gray}{0.5}

\input xy
\xyoption{all}

\begin{document}

\title{Graph model of the Heisenberg-Weyl algebra}

\author{P Blasiak$^1$, A Horzela$^1$, G H E Duchamp$^2$, K A Penson$^3$ \newline and A I Solomon$^{3,4}$\vspace{0.2cm}}

\address{$^1$ H. Niewodnicza\'nski Institute of
Nuclear Physics, Polish Academy of Sciences\newline
ul.\ Radzikowskiego 152, PL 31342 Krak\'ow, Poland\vspace{0.2cm}}

\address{$^2$ Universit\'e Paris-Nord, LIPN, CNRS UMR 7030\newline
99 Av.\ J.-B.\ Clement, F-93430 Villetaneuse, France\vspace{0.2cm}}

\address{$^3$
Universit\'e Pierre et Marie Curie, LPTMC, CNRS UMR 7600\newline
Tour 24 - 2i\`{e}me \'et., 4 pl.\ Jussieu, F 75252 Paris Cedex 05, France\vspace{0.2cm}}

\address{$^4$ The Open University, Physics and Astronomy Department\newline
Milton Keynes MK7 6AA, United Kingdom\vspace{0.2cm}}

\ead{pawel.blasiak@ifj.edu.pl, ghed@lipn-univ.paris13.fr, andrzej.horzela@ifj.edu.pl, penson@lptmc.jussieu.fr, a.i.solomon@open.ac.uk}

\begin{abstract}
We consider an algebraic formulation of Quantum Theory and develop a combinatorial model of the Heisenberg--Weyl algebra structure.
It is shown that by lifting this structure to the richer algebra of graph operator calculus, we  gain a  simple
interpretation involving, for example, the  natural composition of graphs. This provides a deeper insight into the algebraic structure of Quantum Theory
and sheds light on the intrinsic combinatorial underpinning of its abstract formalism.
\end{abstract}

\section{Introduction}\label{Introduction}\vspace{0.2cm}
Quantum Theory seen in action is an interplay of mathematical ideas and physical concepts.
From a present-day perspective its formalism and structure is founded
on the theory of Hilbert space \cite{IshamBook,PeresBook,BallentineBook}.
According to a few basic postulates, the physical notions of  transformations and measurements on a system
are described in terms of operators.
In this way the algebra of operators constitutes the proper
mathematical framework within which quantum theories are built.
The structure of this algebra is determined by two operations, the addition and multiplication of operators;
this lies at the root of all fundamental aspects of Quantum Theory \cite{DiracBook}.

However, the physical content of Quantum Theory transcends the abstract mathematical formalism. It is provided by the correspondence rules assigning operators to physical quantities. This is always an \emph{ad hoc} procedure invoking concrete representations of the operator algebra chosen to best reflect the physical concepts related to the phenomena under investigation.  The most common structure in Quantum Theory is the Heisenberg--Weyl algebra. This describes the algebraic relation between the {\em position} and {\em momentum} operators,  equally  the {\em creation} and {\em annihilation} operators,  which provide our link to the most fundamental physical concepts. Accordingly, we take the Heisenberg--Weyl algebra as the central point of our study.

Interest in combinatorial representations of mathematical entities stems from a wealth
of concrete models they provide. Their convenience comes from simplicity,
which, being based on the elementary notion of enumeration, directly appeals to intuition,
often rendering invaluable interpretations illustrating abstract mathematical constructions \cite{FlajoletBook,BergeronBook,AignerBook}.
This makes the combinatorial perspective particularly attractive in quantum physics, given 
 the latter's  active pursuit of a proper understanding of  fundamental phenomena.

In this paper we develop a combinatorial representation of the operator algebra
of Quantum Theory which is based on the Heisenberg--Weyl algebra. We  recast it in the language of graphs with a simple composition rule and show how, from this perspective, abstract algebraic structures gain an intuitive meaning.
In some respects this draws on the Feynman idea of representing physical processes
as diagrams, familiar as a bookkeeping tool in the perturbation
expansions of quantum  field theory \cite{BjorkenDrell,MattuckBook}. The combinatorial approach, however, has much more to offer
if applied to the overall structure of Quantum Theory seen from the algebraic point of view.
We will show that the process of  lifting to a  more structured algebra of graphs
gives the abstract operator calculus a straightforward interpretation, reflecting 
natural operations on graphs.
This provides an interesting insight into the algebraic counterpart of the theory
and sheds light on the intrinsic combinatorial structures which lie behind its abstract formalism.


\section{Quantum Theory as an Algebra of Operators}\label{QuantumTheory}\vspace{0.2cm}


The usual  setting for Quantum Theory consists of specifying a Hilbert space $\mathsf{H}$, whose vectors are the states of  the system, and identifying operators on those states with physically relevant quantities. Operators acting on $\mathsf{H}$ naturally form an algebra with addition and multiplication, which we denote by $\mathcal{O}$. The most interesting structures in $\mathcal{O}$ are, of course, those generated by operators having a physical interpretation. They usually originate from considering some observables of interest along with operations causing changes in the state of a system.
Accordingly, one takes a hermitian operator, say $N$, representing some observable
and defines a  basis in $\mathsf{H}$
related to states with definite values of the corresponding physical quantity.
The eigenvectors of $N$  are given by $N|n\rangle=n|n\rangle$, numbering the chosen eigenbasis $|n\rangle$ in $\mathsf{H}$ ($n=0,1,2,...$).
One is then interested in describing processes which change the state of the system, \emph{e.g.} time evolution, interactions and other transformations. For that purpose it is convenient to introduce annihilation $a$ and creation $a^\dag$ operators which shift the basis vectors by one, \emph{i.e.}
$[a,N] = a$ and $[a^\dag,N] = - a^\dag$.
Conventionally, these operators are required to satisfy the canonical commutation relation
\begin{eqnarray}\label{HW}
[a,a^\dag]=1\,,
\end{eqnarray}
constituting the Heisenberg--Weyl algebra structure \cite{ArtinBook,BourbakiAlgebraI} which has became the hallmark of noncommutativity in Quantum Theory \cite{DiracBook}.
The operators defined above play the role of elementary processes altering the system by changing its state
with respect to the chosen physical characteristic, \emph{i.e.} they cause a jump between the eigenstates $|n\rangle$ according to the rule $a\,|n\rangle=\sqrt{n}\,|n-1\rangle$ and $a^\dag\,|n\rangle=\sqrt{n+1}\,|n+1\rangle$.
We shall assume that any change of state can be obtained by the action of  some combination of such
creation and annihilation acts, making the operators $a$ and $a^\dag$ convenient building blocks describing the transformations of a system.


The creation and annihilation operators can be used to represent elements of the algebra $\mathcal{O}$.
Indeed, each operator can be seen as an element of the free algebra generated by $a$ and $a^\dag$,  \emph{i.e.} written as a linear combination of words in generators.
This procedure is, however, ambiguous due to the commutation relation Eq.~(\ref{HW})
which yields different representations of the same operator \cite{CahillGlauber}.
To solve this problem, the order of $a$ and $a^\dag$ has to be fixed.
Conventionally this is done by choosing the normally ordered form in which
all annihilators stand to the right of creators \cite{KlauderBook,AmJPhys}.
Consequently, each operator $A\in\mathcal{O}$ can be uniquely written in the normally ordered form such as\footnote{We do not specify limits of summation and constraints on coefficients
since it does not affect the algebraic considerations and can be introduced at each step if needed.}
\begin{eqnarray}\label{A}
A=\sum_{r,s}\alpha_{rs}\ a^{\dag\,r} a^s.
\end{eqnarray}
In this way elements of the operator algebra $\mathcal{O}$ are represented in terms of the {\em ladder} operators
$a$ and $a^\dag$, and interpreted as combinations of the elementary acts of annihilation and creation.
Eq.~(\ref{A}) will be the starting point of our combinatorial representation of the algebra $\mathcal{O}$.

\section{Graphs and their Algebra}\label{Graphs}\vspace{0.2cm}

 Considering combinatorial realizations of operator algebras,  we shall specify
two classes of graphs $\mathfrak{g}$ and $\mathfrak{g_1}$,
the latter being the shadow of the former under a suitable {\em forgetful} procedure.
We shall employ the convenient notion of graph composition to show how these structures
are naturally made into algebras, providing the representation of the algebra $\mathcal{O}$.

\subsection{Graphs $\&$ Composition}\vspace{0.2cm}

A graph is a collection of vertices connected by lines with internal structure determined by some construction rules. For our  purposes,  we consider a specific class of graphs defined in the following way.

\subsubsection{{Vertices $\&$ Lines.}}
\begin{figure}[htp]
\begin{center}
\resizebox{0.4\columnwidth}{!}{\includegraphics{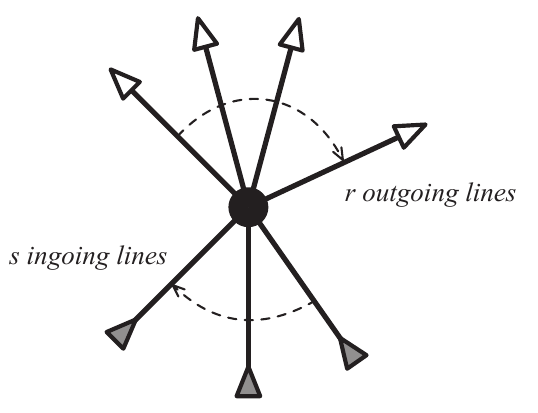}}
\caption{\label{Fig1} A generic one-vertex graph $\varGamma^{(r,s)}\in\mathfrak{g_1}$. }
\end{center}
\end{figure}

The basic building blocks of the graphs are vertices $\bullet$ attached to two sorts of lines,
those coming into, and those going out of,  the vertex, having loose ends marked with grey ${{\color{light-gray}\blacktriangle}\!\!\!\!\!\!\vartriangle}$ and white $\vartriangle$ arrows respectively.
A generic one-vertex graph $\varGamma^{(r,s)}$ is characterized by two numbers $r$ and $s$
counting incoming and outgoing lines respectively, see Fig.~\ref{Fig1}.
 We shall denote by $\mathfrak{g_1}$  the class comprising  all such one-vertex graphs, and by \O\ the empty, or void, graph  (no vertices, no lines).
In a further construction we shall assume that all lines attached to vertices are distinguishable.

\subsubsection{{Construction Rules.}}
\begin{figure}[htp]
\begin{center}
\resizebox{0.7\columnwidth}{!}{\includegraphics{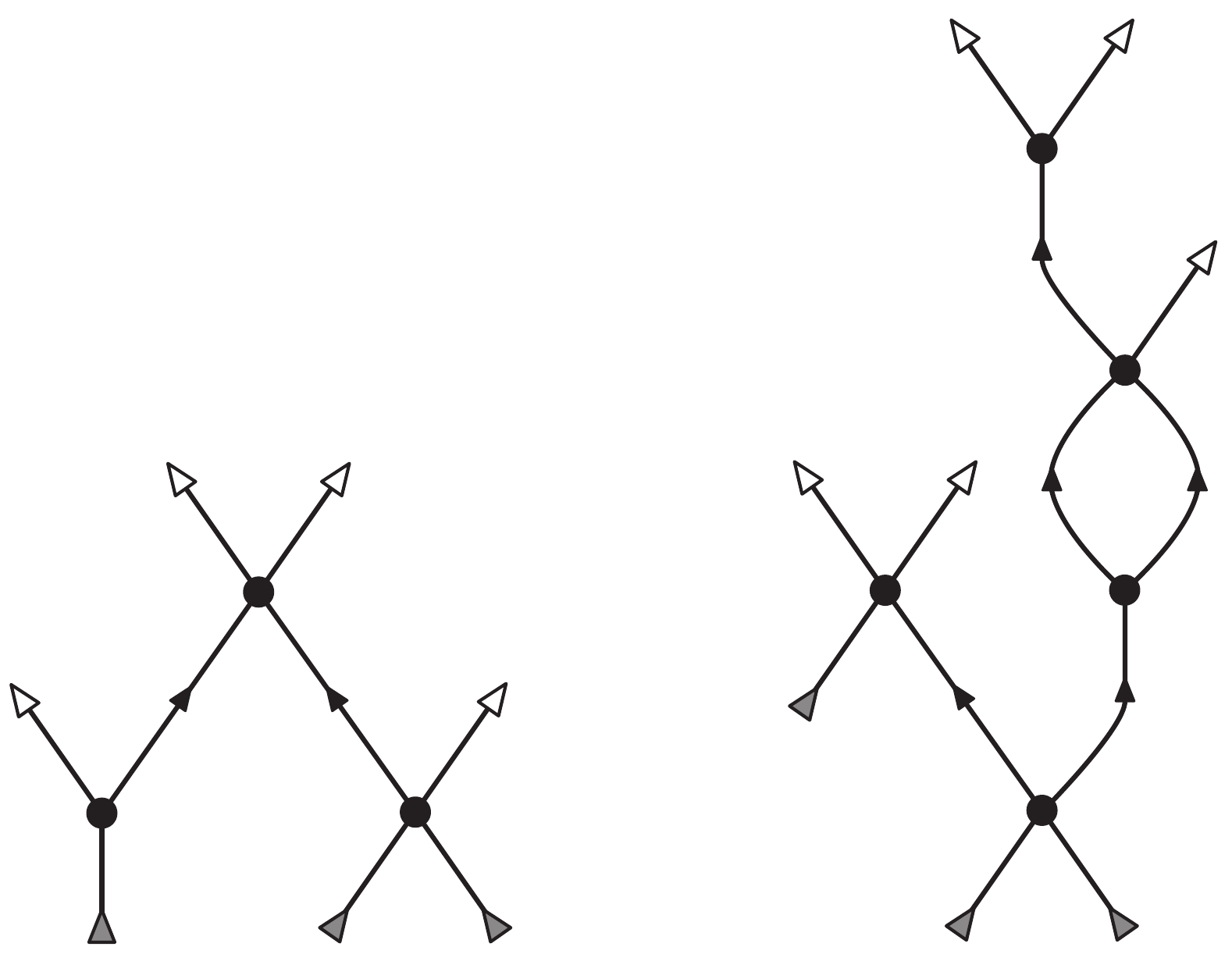}}
\caption{\label{Fig2} Example of a multi-vertex graph (8 vertices and 9 outgoing lines, 6 incoming lines, 5 inner lines) built of two kinds of vertices: $\varGamma^{(2,2)}$ ($X$ shape) and $\varGamma^{(2,1)}$ ($Y$ shape).}
\end{center}
\end{figure}

A multi-vertex graph $\varGamma$ is a set of vertices with additional structure introduced by
joining some of the outgoing lines to the incoming ones.
The requirement that the original direction of lines is preserved results in a {\em directed} structure of graphs
indicated by black arrows $\blacktriangle$ on the inner lines. We further restrict the class of  graphs we consider 
to those without cycles, \emph{i.e.} we exclude graphs with closed paths.
An example of a multi-vertex graph is shown in Fig.~\ref{Fig2}.

The rules specified above define the class of graphs denoted by $\mathfrak{g}$.
In a less formal manner, we can describe these graphs as having an inner structure determined
by directed connections between vertices and a characteristic set of outer lines
marked with grey ${{\color{light-gray}\blacktriangle}\!\!\!\!\!\!\vartriangle}$ and white
$\vartriangle$ arrows at the loose ends.
These graphs can be seen as a kind of process,
with the vertices being intermediate steps.
This observation can be developed further with the help of the convenient notion of {\em graph composition}.

\begin{figure*}[h]
\begin{center}
\resizebox{\columnwidth}{!}{\includegraphics{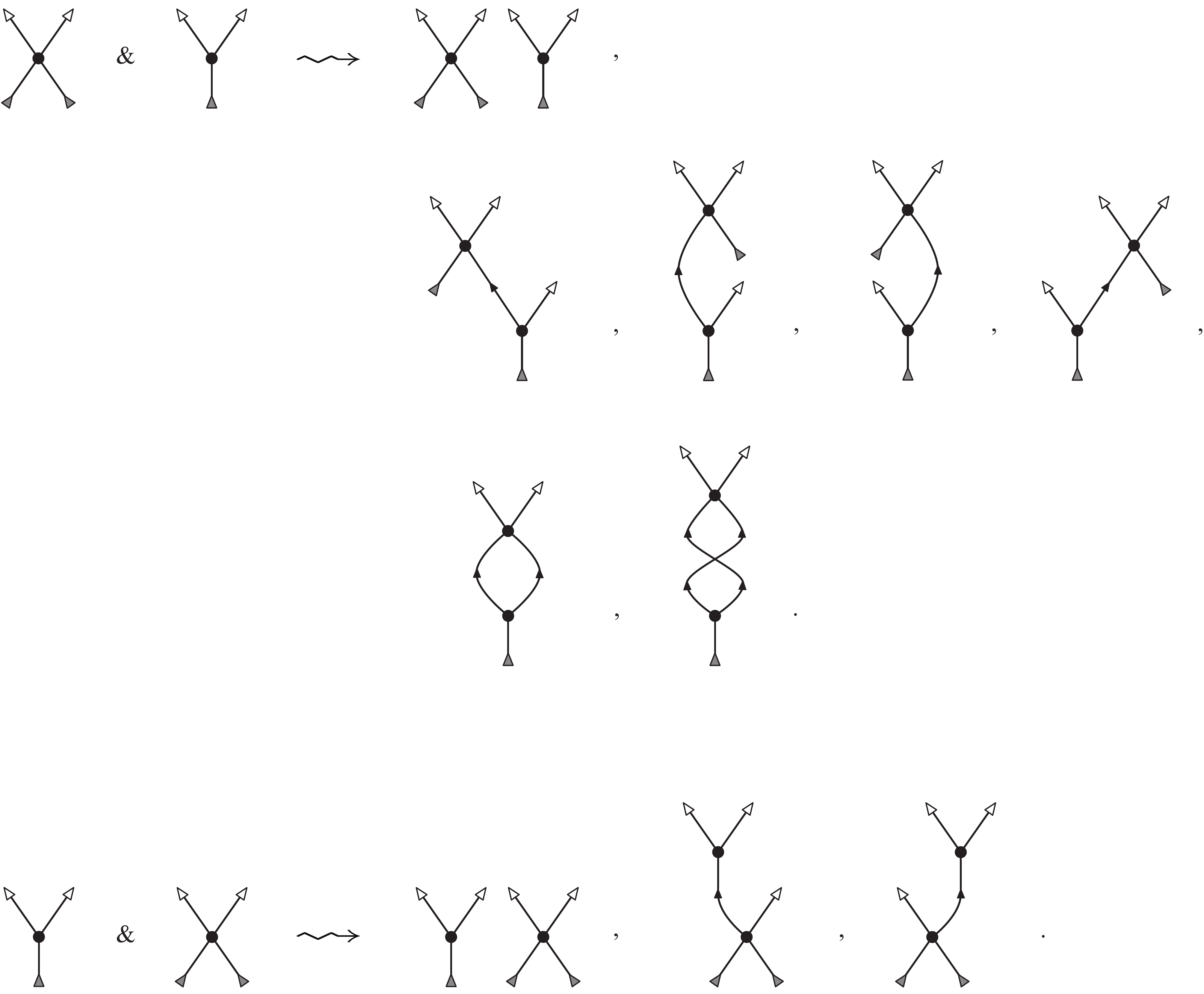}}
\end{center}
\caption{\label{Fig3}Two one-vertex graphs composed in different order.
Note that distinguishability of the lines is taken into account.}
\end{figure*}

\subsubsection{{Composition.}} Two graphs can be composed by
joining some of the incoming lines (grey arrows) of the first one
with some of the outgoing lines (white arrows) of the second one.
This operation is {\em inner} in $\mathfrak{g}$ since it preserves the direction of the lines
and does not introduce cycles. Observe that two graphs can be composed in many ways,
\emph{i.e.} as many as there are possible choices of pairs of lines
(grey arrows from the first one and white arrows from the second one)
which are joined, see Fig.~\ref{Fig2}.
Note also that composing graphs in  reverse order yields different results.



The notion of graph composition allows for an iterative definition,
\emph{i.e.} any element of $\mathfrak{g}$ can be constructed
starting from the void graph by successive composition
with one-vertex graphs.
Consequently, the class of one-vertex graphs $\mathfrak{g_1}\subset\mathfrak{g}$
can be seen as representing basic processes or events happening one after another
and constituting  a composite process -- a multi-vertex graph.

\subsection{Equivalence of Graphs}\vspace{0.2cm}

In many cases one is not interested in the inner structure of a graph
and needs only  to focus on the outer lines.
This is equivalent to considering the graph's one-vertex equivalents obtained
by replacing all inner vertices and lines by a single vertex
and keeping all the outer lines untouched, \emph{i.e.} $\varGamma\stackrel{_\thicksim}{\longrightarrow}\varGamma^{(r,s)}$
where $\varGamma$ is the graph with $r$ outgoing and $s$ incoming lines.
For example, for the graph in  Fig.~\ref{Fig2} this gives $\varGamma\stackrel{_\thicksim}{\longrightarrow}\varGamma^{(9,6)}$.
The mapping $\mathfrak{g}\stackrel{_\thicksim}{\longrightarrow}\mathfrak{g_1}$ reduces to
forgetting about the inner structure of graphs and introduces an equivalence relation in $\mathfrak{g}$.
Accordingly, two graphs are {\em equivalent} $\varGamma_1\thicksim\varGamma_2$
if and only if both have the same number of incoming and outgoing lines respectively.
The simplest choice of representatives of equivalence classes
are the one-vertex graphs and so the quotient set $\mathfrak{g}/_{\!\thicksim}$
is isomorphic to the set of one-vertex graphs $\mathfrak{g_1}$.

There are two characteristic mappings between $\mathfrak{g}$ and $\mathfrak{g_1}$: the canonical projection map described above and the inclusion map $\mathfrak{g_1}\subset\mathfrak{g}$, \emph{i.e.}
\begin{eqnarray}\label{Diag_g}
	\xymatrix{
	\mathfrak{g}\ \ar@/^/[rr]^{\thicksim}&&\ar@/^/[ll]^{\supset}\ \mathfrak{g_1}
	}
\end{eqnarray}
In this sense $\mathfrak{g_1}$ is a shadow of the more structured class $\mathfrak{g}$.
Observe that the arrows in Diag.~(\ref{Diag_g}) can not be reversed,
\emph{i.e.} once the inner structure of a graph is forgotten it cannot be restored.

\subsection{Algebra of Graphs}\vspace{0.2cm}

Both classes of graphs $\mathfrak{g}$ and $\mathfrak{g_1}$ can be endowed
with the structure of a noncommutative algebra based on the natural concept of graph composition.
An algebra requires two operations, addition and multiplication, which are constructed as follows.
We define the vector space $\mathcal{G}$ (over $\mathbb{C}$) generated by the basis set $\mathfrak{g}$, by
\begin{eqnarray}\label{G}
\mathcal{G}=\left\{\ \left.\sum\right._i\alpha_i\ \varGamma_i:\ \alpha_i\in\mathbb{C},\ \varGamma_i\in\mathfrak{g}\ \right\}.
\end{eqnarray}
Addition in $\mathcal{G}$ has the usual form
\begin{eqnarray}\label{Addition}
{\sum}_i\ \alpha_i\ \varGamma_i+{\sum}_i\ \beta_i\ \varGamma_i={\sum}_i\ (\alpha_i+\beta_i)\ \varGamma_i.
\end{eqnarray}
The less trivial  part in the definition of an algebra $\mathcal{G}$ concerns multiplication, which by bilinearity
\begin{eqnarray}
{\sum}_i\ \alpha_i\ \varGamma_i\,*\,{\sum}_j\ \beta_j\ \varGamma_j={\sum}_{i,j}\ \alpha_i\beta_j\ \varGamma_i\,*\,\varGamma_j,
\end{eqnarray}
only requires definition on the basis set $\mathfrak{g}$.
Recalling the notion of graph composition,  the definition which suggests itself is (compare with Fig.~\ref{Fig3})
\begin{eqnarray}\label{GG}
\varGamma_i\,*\,\varGamma_j=\sum\ all\ compositions\ of\ \varGamma_i\ with\ \varGamma_j\ .
\end{eqnarray}
Note that all the terms in the sum are distinct with coefficient equal to one.
The multiplication thus defined is noncommutative and produces $\mathcal{G}$
an associative algebra with unit (void graph).

Imposing an algebraic structure on $\mathfrak{g_1}$ roughly follows the above scheme.
Accordingly, one defines the vector space
\begin{eqnarray}\label{G1}
\mathcal{G}_{\mathfrak{1}}=\left\{\ \sum\right._{i,j}\alpha_{i,j}\ \varGamma^{(i,j)}:\ \alpha_{i,j}\in\mathbb{C},\ \left.\varGamma^{(i,j)}\in\mathfrak{g_1}\ \right\},
\end{eqnarray}
with addition defined analogously to Eq.~(\ref{Addition}).
Multiplication again reduces to defining it on the basis set $\mathfrak{g_1}$,
but the  obstacle here is that the composition rule is not closed  within the class $\mathfrak{g_1}$,
\emph{i.e.} it produces  two-vertex graphs which belong to $\mathfrak{g}$.
This however can be overcome by applying the forgetful mapping $\varGamma\stackrel{_\thicksim}{\longrightarrow}\varGamma^{(r,s)}$ to the result.
Specifically, multiplication of two graphs in $\mathfrak{g_1}$ follows  Diag.~(\ref{Diag_g})
and consists of:
\begin{enumerate}
\item[1)]{treating graphs as elements of $\mathfrak{g}$,}
\item[2)]{multiplying them according to Eq.~(\ref{GG}),}
\item[3)]{forgetting the inner structure of the resulting two-vertex graphs in the sum.}
\end{enumerate}
Note that, contrary to Eq.~(\ref{GG}), some of the resulting terms may be  equal and their sum may involve
nontrivial integer coefficients.
Grouping terms with respect to the number of joined lines yields the explicit formula
\begin{eqnarray}\label{GGG}
\varGamma^{(r,s)}\,*\,\varGamma^{(k,l)}=\sum_{i=0}^{min\{k,s\}}\binom{s}{i}\binom{k}{i}i!\ \ \varGamma^{(r+k-i,s+l-i)}.
\end{eqnarray}
For example: $\varGamma^{(2,1)}\,*\,\varGamma^{(2,2)}=\varGamma^{(4,3)}+2\,\varGamma^{(3,2)}$
and $\varGamma^{(2,2)}\,*\,\varGamma^{(2,1)}=\varGamma^{(4,3)}+4\,\varGamma^{(3,2)}+2\,\varGamma^{(2,1)}$,
see Fig.~\ref{Fig3}. In this way, multiplication in the richer structure $\mathcal{G}$
is naturally projected onto $\mathcal{G}_{\mathfrak{1}}$.
The resulting  combinatorial algebra $\mathcal{G}_{\mathfrak{1}}$ is associative and noncommutative.
Again, similarly to $\mathfrak{g}$ and $\mathfrak{g_1}$ in Diag.~(\ref{Diag_g}),
both algebras are related by
\begin{eqnarray}\label{Diag_G}
	\xymatrix{
	\mathcal{G}\ \ar@/^/[rr]^{\thicksim}&&\ar@/^/[ll]^{\supset}\ \mathcal{G}_{\mathfrak{1}}
	}
\end{eqnarray}
Hence, the algebra $\mathcal{G}_{\mathfrak{1}}$ is an image of the more structured algebra of graphs $\mathcal{G}$.

\section{Graph representation of operator algebra}\label{OperatorsGraphs}\vspace{0.2cm}

The structures described above are  examples of algebras having concrete representations
based on the natural concept of graph composition.
It appears that both are intimately related to the algebra of operators $\mathcal{O}$.
As suggested by the similarity of elements in $\mathcal{O}$ and $\mathcal{G}_{\mathfrak{1}}$,
see Eqs.~(\ref{A}) and (\ref{G1}), we make a correspondence of the basis sets
\begin{eqnarray}\label{AG}
a^{\dag\,r} a^s\ \longleftrightarrow\ \varGamma^{(r,s)}
\end{eqnarray}
establishing the isomorphism of the vector spaces.
This  would not be  surprising if it were not for  the fact that this mapping also preserves multiplication in both algebras.
Indeed, multiplication of the basis elements in $\mathcal{O}$ gives
\begin{eqnarray}\label{AAA}
a^{\dag\,r}a^s\, a^{\dag\,k}a^l=\!\!\sum_{i=0}^{min\{k,s\}}\!\!\binom{s}{i}\binom{k}{i}i!\ \ a^{\dag\,{r+k-i}}a^{s+l-i},
\end{eqnarray}
which is the result of transforming $a^sa^{\dag\,k}$
to the normally ordered form using the commutator
$[a^s,a^{\dag\,k}]=\sum_{i=1}^{min\{k,s\}}\binom{s}{i}\binom{k}{i}i!\ a^{\dag\,k-i}a^{s-i}$, see \emph{e.g.} \cite{BlasiakJPA2008}.
It suffices to compare Eqs.~(\ref{GGG}) and (\ref{AAA}) to show that Eq.~(\ref{AG})
establishes an isomorphism between the algebras $\mathcal{O}$ and $\mathcal{G}_{\mathfrak{1}}$.
In other words, both algebras are essentially the same,
\emph{i.e.} they have all elements and operations equivalent.
We can thus enjoy the advantages of concrete realization
of the abstract operator algebra in terms of graphs.
For example, instead of multiplying operators in $\mathcal{O}$
one can do it in $\mathcal{G}_{\mathfrak{1}}$ simply by composing graphs.
The crucial role in this procedure is played by the more structured algebra of graphs $\mathcal{G}$
where all these operations have a simple interpretation.
Accordingly, Diag.~(\ref{Diag_G}) can be complemented to
\begin{eqnarray}\label{GGO}
	\xymatrix{
	\mathcal{G}\ \ar@/^/[rr]^{\thicksim}&&\ar@/^/[ll]^{\supset}\ \mathcal{G}_{\mathfrak{1}}\ \ar@{<->}[rr]^{\ \mathit{1:1}}&&\ \mathcal{O}
	}
\end{eqnarray}
In this way, the algebra $\mathcal{O}$ gains a combinatorial representation via $\mathcal{G}_{\mathfrak{1}}$
and can be seen as reflecting the natural processes taking place in $\mathcal{G}$.

\section{Discussion}\label{Conclusion}\vspace{0.2cm}

We have considered the Heisenberg--Weyl algebra within the operator structure of Quantum Theory
and exploited a convenient representation of operators
by normally ordered expressions in ladder operators $a$ and $a^\dag$.
This allowed us to  identify the operator algebra $\mathcal{O}$
with the combinatorial algebra $\mathcal{G}_{\mathfrak{1}}$
constructed as the projection of the algebra of graphs $\mathcal{G}$.
The main result of the paper shows that these combinatorial structures
provide us with a simple interpretation of the operator calculus
in terms of the natural composition of graphs.
Accordingly, the operator algebra can be seen as an image (via $\mathcal{G}_{\mathfrak{1}}$)
of the algebra $\mathcal{G}$ which is a categorical  version of the algebra $\mathcal{O}$ \cite{Baez,Morton}.
This may be illustrated  by redrawing Diag.~(\ref{GGO}) as\vspace{0.3cm}
\begin{eqnarray}\nonumber
	\xymatrix{&\ \mathcal{G}\ \ar[dr]^{\!\!\!\rotatebox[origin=r]{-45}{$\thicksim$}}&&&\ \ \ \ \ \textsc{Graph\ Algebra}\ \ \ \ar@{~>}[l]\\
	\ \mathcal{O}\ \ar[ur]^{\rotatebox[origin=c]{45}{$\subset$}\!\!\!}\ar@{<->}[rr]^{\mathit{1:1}}&&\ \mathcal{G}_{\mathfrak{1}}\ &&\ \ \ \substack{{\displaystyle{\textsc{Quantum\ Theory}}}\\Heisenberg-Weyl\ algebra}\ar@{~>}[l]
	}
\end{eqnarray}
The diagram indicates the existence of the fundamental graph structure $\mathcal{G}$
of which the algebraic structure of Quantum Theory is just the reflection.
That is, all objects in the theory, as well as its calculus,  can be seen, via simple inclusion,
as elements and operations present in the richer algebra of graphs
described by natural composition rules.
At any time it is possible to return to the coarser level of an algebra of operators,
simply by  forgetting about the inner structure of graphs.
As a result, the algebra of graphs suggests itself as a fundamental combinatorial level of Quantum Theory.
The  process described here, of  lifting the theory to the richer structure $\mathcal{G}$,
is motivated by the natural interpretation of graphs as processes transforming
quantities or objects, which is an attractive concept from the physical point of view.
Moreover, the major advantage of the combinatorial representation of the algebra $\mathcal{O}$,
presented above, is that the abstract operator calculus can be seen intuitively as
a straightforward composition of graphs.



\ack\vspace{0.2cm}

We wish to thank Philippe Flajolet for important discussions on the subject.
Most of this research was carried out in the Mathematisches Forschungsinstitut Oberwolfach (Germany) and the Laboratoire d'Informatique de l'Universit\'e Paris-Nord in Villetaneuse (France) whose warm hospitality is greatly appreciated.
The authors acknowledge support from the Polish Ministry of Science and Higher Education grant no. N202 061434 and the Agence Nationale de la Recherche under programme no. ANR-08-BLAN-0243-2.

\section*{References}\vspace{0.2cm}

\bibliography{Bibliography}
\bibliographystyle{iopart-num}

\end{document}